\RequirePackage{fix-cm}
\documentclass[smallextended]{svjour3}       % onecolumn (second format)
\smartqed  % flush right qed marks, e.g. at end of proof
\usepackage{graphicx}
\usepackage{mathptmx}          
\begin{document}

\title{Review \\ Self-ordered nanostructures on patterned substrates%\thanks{Grants or other notes
}
\subtitle{Experiment and theory of metalorganic vapor-phase epitaxy of V-groove quantum wires and pyramidal quantum dots}


\author{Emanuele Pelucchi \and Stefano T. Moroni \and Valeria Dimastrodonato \and Dimitri D. Vvedensky}


\institute{Emanuele Pelucchi, Stefano T. Moroni, and Valeria Dimastrodonato \at 
Tyndall National Institute, University College Cork, ``Lee Maltings,'' Dyke Parade, Cork, Ireland \and
Dimitri D. Vvedensky \at
The Blackett Laboratory, Imperial College London, London SW7 2AZ, United Kingdom}

\date{Received: date / Accepted: date}

\maketitle

\begin{abstract}
The formation of nanostructures during metalorganic vapor-phase epitaxy on patterned (001)/(111)B GaAs substrates is reviewed. The focus of this review is on the seminal experiments that revealed the key kinetic processes during nanostructure formation and the theory and modelling that explained the phenomenology in successively greater detail. Experiments have demonstrated that V-groove quantum wires and pyramidal quantum dots result from self-limiting concentration profiles that develop at the bottom of V-grooves and inverted pyramids, respectively.  In the 1950s, long before the practical importance of patterned substrates became evident, the mechanisms of capillarity during the equilibration of non-planar surfaces were identified and characterized.  This was followed, from the late 1980s by the identification of growth rate anisotropies (i.e.~differential growth rates of crystallographic facets) and precursor decomposition anisotropies, with parallel developments in the fabrication  of V-groove quantum wires and pyramidal quantum dots.  The modelling of these growth processes began at the scale of facets and culminated in systems of coupled reaction-diffusion equations, one for each crystallographic facet that defines the pattern, which takes account of the decomposition and surface diffusion kinetics of the group-III precursors and the subsequent surface diffusion and incorporation of the group-III atoms released by these precursors.   Solutions of the equations with optimized parameters produced concentration profiles that provided a quantitative interpretation of the time-, temperature-, and alloy-concentration dependence of the self-ordering process seen in experiments. 

\keywords{quantum wires \and quantum dots \and epitaxy \and self-limiting \and metalorganic precursors \and vapor-phase \and theory \and reaction-diffusion equations \and precursor decomposition \and surface diffusion}
\PACS{68.55.-a \and 68.65.-k \and 81.05.Ea \and 81.10.A}
\end{abstract}

%%%%%%%%%%%%%%%%%%%%%%%%%%%%%%%%%%%%%%%%%%%%%%%%%%

\section{Introduction}
\label{sec1}

The lithographic patterning of a surface exposes crystallographic facets with different chemical, transport, and structural properties. Accompanying such variations of individual facet properties is an interaction mediated by interfacet mass transfer. These factors conspire to produce a nonuniform growth rate across a patterned substrate during the deposition of new material that can be exploited to influence the position, size, and composition of nanostructures, such as quantum wires and quantum dots.  

Metalorganic vapor-phase epitaxy (MOVPE) \cite{stringfellow99,jensen93}, based on the hydrodynamic delivery of the atomic constituents of the nanostructure within polyatomic molecules called {\it precursors} is particularly well-suited to the foregoing scenario.  The crystallographic orientation of the substrate and the exposed facets can be chosen to ensure that growth occurs predominantly within the etched patterns, which act as a template for a particular nanostructure. 

MOVPE on patterned substrates has enabled the development of high-quality ordered semiconductor nanostructures for applications to optoelectronic and integrated quantum optics \cite{kapon89,madhukar93,koshiba93,notzel98,kapon99}. The initial pattern evolves toward a stationary shape resulting from the interplay between the pattern and the facet-dependent kinetics \cite{ozdemir92,sato04} determined by the growth conditions (growth rate, temperature, and material composition). This enables the fabrication of morphologically controlled nanostructures with electro-optical features that can be tuned by the pattern and growth conditions.

A wide variety of patterns has been used in the ongoing effort to produce uniform arrays of nanostructures (usually quantum dots) at specified positions, including triangles \cite{fukui91}, squares \cite{kumakura95}, wires and dots \cite{labena90}, striped and square mesas \cite{araki97}, truncated triangular pyramidal mesas \cite{madhukar93}, patterned high-index substrates \cite{notzel98}, V-grooves \cite{kapon89}, and inverted pyramids \cite{sugiyama95,tsujikawa97}. In this review we will focus on nanostructures formed near the bases of V-grooves (quantum wires) and inverted pyramids (quantum dots) because of the wealth of systematic measurements made on their growth, optical properties, and supporting theory and modelling.

The organization of this review is as follows. The patterned substrates that are used to form V-groove quantum wires and pyramidal quantum dots are summarized in Sec.~\ref{sec2}, together with the seminal experimental work that reported these structures and revealed their fundamental properties.   The development of the current understanding of the fundamental principles of nanostructure formation on patterned substrates is the subject of Secs.~\ref{sec3} and \ref{sec4}, with early work discussed in Sec.~\ref{sec3} and more recent developments in Sec.~\ref{sec4}.  We provide an outlook for future work in Sec.~\ref{sec5}.

%%%%%%%%%%%%%%%%%%%%%%%%%%%%%%%%%%%%%%%%%%%%%%%%%%

\section{V-Grooves and Inverted Pyramids}
\label{sec2}

\begin{figure}
\includegraphics[width=0.6\textwidth]{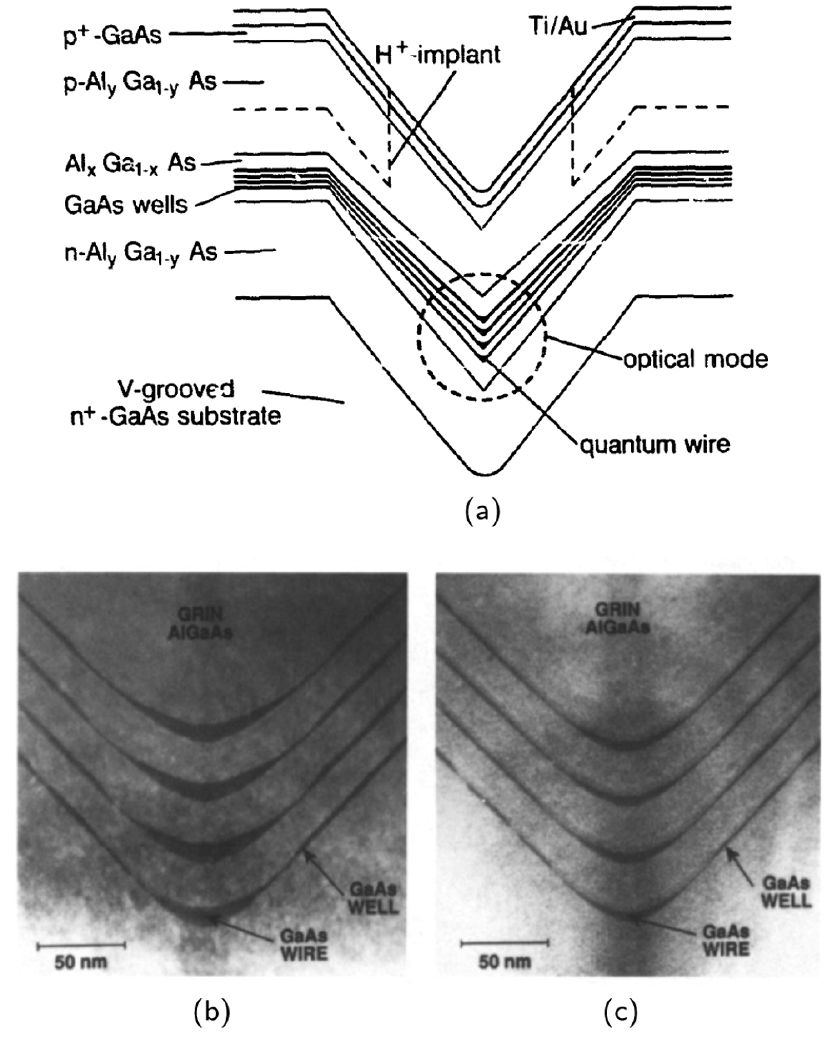}
\caption{(a) Schematic representation of a V-groove 4-quantum-wire GaAs/AlGaAs laser. (b,c)Transmission electron microscopy  cross-sectional view of the cores of two lasers with different wire size grown by MOVPE \cite{kapon99,kapon92}. Reprinted from Surface Science, \textbf{267}, E. Kapon, D.M. Hwang, M. Walther, R. Bhat, N.G. Stoffel, \textit{Two-dimensional quantum confinement in multiple quantum wire lasers grown by OMCVD on V-grooved substrates}, pp 593-600, Copyright (1992), with permission from Elsevier.}
\label{fig1}       
\end{figure}

The term ``V-groove'' is derived from the geometry of the recesses in which epitaxial growth is performed, that is, a semiconductor wafer (typically (100) GaAs) which has been patterned with conventional lithography to obtain a groove along the $(0 1\bar{1})$ direction). The cross-section typically shows a V-shape, originating from the simplest of the methodologies used to obtain the recess by wet chemical etching. Indeed, as is well known, a number of chemical solutions is selective to the relatively sturdy (111)A surfaces in GaAs, i.e.~for a ``striped'' opening in a resist mask on top of a (100) wafer, the etchant would consume the GaAs and cease (or substantially  diminish, to be precise) when encountering the (111)A surfaces, producing the substrate depicted in Fig.~\ref{fig1}.

Some important milestones have been achieved:~the formation of (single and multiple) quantum wires at the center of the recesses (Fig.~\ref{fig1}(b,c)), and (low) threshold lasing obtained already in the late 1980s and in later years \cite{kapon99}. Nevertheless such quantum-wire lasers never met the requirements for industrial exploitation, and most of the subsequent research concentrated on the physics of one-dimensional systems, which yielded some important insights (see, for example, \cite{vouilloz92,feltrin97,levy06}). In a way, V-groove quantum wires can be considered forerunners of the broad modern effort dedicated to nanowires obtained by vapor-liquid-solid methods and variations thereon, and their properties. (e.g.~\cite{law04,lu08})

\begin{figure}
\includegraphics[width=\textwidth]{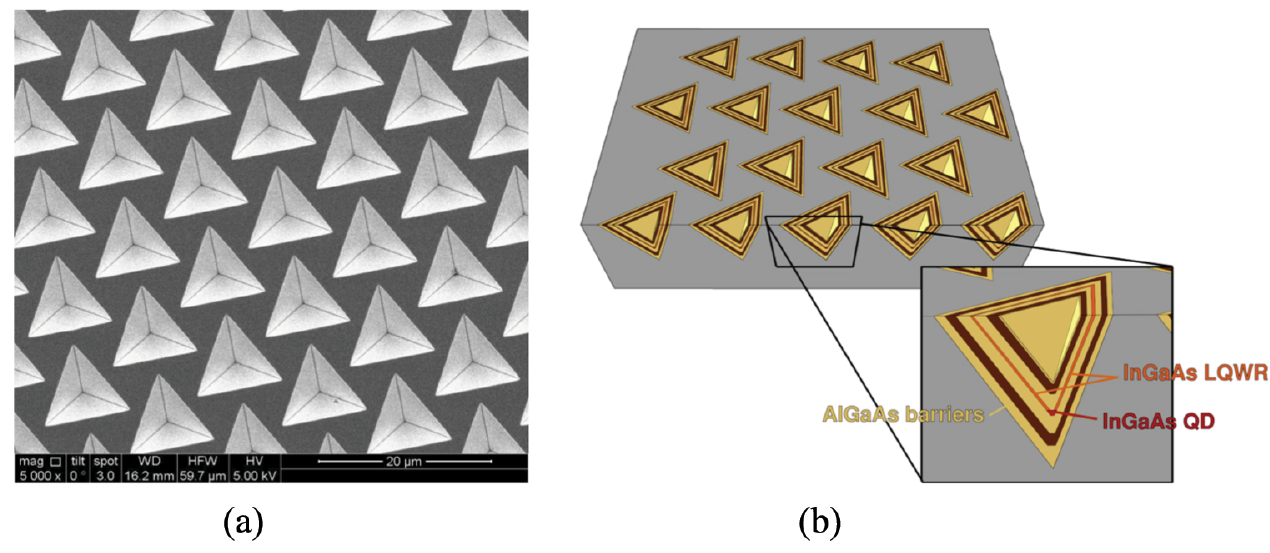}
\caption{(a) An array of 10~$\mu$m-spaced as-etched pyramidal recesses on a (111)B GaAs surface. (b) Schematic representation of an as-grown array of pyramidal quantum dots. The inset shows a quantum dot, lateral quantum wires, and AlGaAs barriers.}
\label{fig2}       
\end{figure}

An important off-shoot of the V-groove research has been driven by the possibility of obtaining zero-dimensional nanostructures by epitaxial growth on pre-patterned substrates, but shifting from two-dimensional V-grooves to three-dimensional recesses. Indeed, if a circular and/or triangular pattern is opened in a resist/SiO$_2$ mask (or similar) deposited onto a (111)B GaAs substrate, (selective) wet etching will again stop on the three (111)A surfaces and provide an inverted pyramidal recess (Fig.~\ref{fig2}(a)). The same processes that yield quantum wires near the bottom of V-grooves, when performed on a (111)A GaAs substrate with an inverted pyramidal pattern, provide single quantum dots at the apex of the inverted recess (Fig.~\ref{fig2}(b)).

This method was pioneered on GaAs by Japanese groups \cite{sugiyama95,tsujikawa97}, but the most significant developments came shortly thereafter from Kapon's group, who obtained several important results, including the unequivocal demonstration of single-dot properties of ``pyramidal'' quantum dots (``pyramidal'' refers to the shape of the recess, not the actual dot shape) \cite{hartmann97,hartmann98,hartmann00}. This became relevant as interest in quantum dots had already begun to shift from commercial applications (e.g.~low-threshold lasers), to more fundamental research in the emerging area of quantum information processing. Indeed, in 2000, single-photon emission was demonstrated in Stranski--Krastanov quantum dots  \cite{michler00,santon01}, and, in 2006, entangled photon emission from ``similar'' quantum dots \cite{akopian00,youmg08}, establishing a major milestone by confirming the ``artificial atom'' picture and its relevance for quantum information processing.  Single-photon emission was shown for single site-controlled pyramidal quantum dots in 2004 \cite{baier04} and entangled-photon emission in 2013 \cite{juska13,juska15}, matching their self-assembled counterparts. 

Site control is a key aspect of quantum dot formation on patterned substrates, and especially so when these structures are scaled up from single demonstrators to active elements of quantum processing devices, which require tens or hundreds of emitters. Pyramidal quantum dots currently have a leading role when excellent optical properties matched to site control are required though, as summarized in the introduction, there are other site-controlled quantum-dot systems.

%%%%%%%%%%%%%%%%%%%%%%%%%%%%%%%%%%%%%%%%%%%%%%%%%%

\section{Early Observations and Models}
\label{sec3}

The initial reports of V-groove quantum wires and pyramidal quantum dots were based entirely on {\it ex situ} measurements, with evidence for the formation of these structures provided directly by transmission electron microscopy (TEM) (Fig.~\ref{fig1}(b,c)), and indirectly by measurements of their emitted light.  An accompanying understanding  of the formation processes of these was still some years away.  For molecular-beam epitaxy (MBE), the availability of {\it in situ} probes, such as reflection high-energy electron diffraction (RHEED) \cite{neave95} and high-temperature scanning tunnelling microscopy \cite{bert01}, fostered a well-developed theory of the rate-determining atomic-scale processes under ultra-high-vacuum conditions.  In contrast, the reactive high-pressure environment during vapor-phase epitaxy limits {\it in situ} measurements to optical methods \cite{fuoss95,aspnes98}, with only X-rays (from a synchrotron) providing information comparable to that of RHEED \cite{fuoss95}.  

In the absence of {\it in situ} measurements for patterned substrates, even the basic observation that V-groove quantum wires and pyramidal quantum dots were formed only by using MOVPE (or metalorganic chemical vapor deposition (MOCVD)) was not fully appreciated, nor was the facet-dependent growth rates resulting from facet-dependent precursor decomposition rates.  MBE  delivered different structures under quite different conditions.  For example, the ``record'' laser obtained in \cite{tiwari94} (see also \cite{kapon99}) used a template morphology substantially different from those  typical used in MOVPE, with the structure obtained because the lithographic mask was maintained on the patterned substrate, covering the ridges and leaving only a fraction of the V-groove exposed to the molecular beam. In MOVPE, the lithographic mask is routinely removed, as there is no need to inhibit or reduce ridge growth.

\subsection{Flattening of a Patterned Substrate}
\label{sec3.1}

When patterning exposes several crystallographic facets, there results a spatially-dependent chemical potential.  This can be seen from the average chemical potential $\mu_0$ just beneath the surface of a facet bounded by $N$ other facets \cite{herring51}:
\begin{equation}
\mu_i=\mu_0+{\Omega\over A_i}\sum_{j=1}^N(\gamma_j\,\mbox{csc}\,\theta_{ij}-\gamma_i\,\mbox{cot}\,\theta_{ij})\ell_{ij}\, ,
\label{eq1}
\end{equation}
where $\mu_0$ is the chemical potential of the bulk crystal, $\Omega$ the atomic volume, $\gamma_i$ is the surface energy of the facet in question with area $A_i$, $\theta_{ij}$ the acute dihedral angle between the $j$th facet, whose surface energy is $\gamma_j$, and the facet in question, and $\ell_{ij}$ the length of the straight boundary between facets $i$ and $j$.  The spatially varying chemical potential indicates that a patterned substrate is not an equilibrium structure.  The resulting thermodynamic driving force, through capillarity, drives mass transfer from regions of high to regions of low chemical potential.  In atomistic terms, thus current  results from the net detachment of atoms from convex sites and a concomitant net attachment near concave sites, where there are more favorable bonding configurations (i.e.~with higher coordination).  In the absence of a deposition flux, this process continues until the surface has planarized and the chemical potential is constant across the surface.  During growth, capillarity {\it competes} with growth, with the resulting morphology depending on the growth conditions.

The morphological evolution of a patterned substrate in the absence of growth was studied by Geguzin and Ovcharenko \cite{geguzin62} long before such substrates were recognized as a route to producing site-controlled nanostructures.   These authors were concerned with the thermal planarization of a patterned substrate by mass transport. For surfaces patterned with either a flat-bottomed groove or a V-groove, they obtained surface profiles of the morphological evolution of the substrate  through either surface diffusion or evaporation/recondensation.  Their predictions for the latter mechanism were confirmed by high-temperature experiments.

Several years earlier, Mullins \cite{mullins57,mullins59,mullins01} considered the effect of capillarity on the shape changes of periodic and other surface profiles.  Mullins identified three contributions to capillarity-induced mass current: surface diffusion, volume diffusion, and evaporation/condensation.  With the intention of obtaining an analytic theory, Mullins invokes two assumptions:~(i) the surface free energy is isotropic, and (ii) the absence of steep slopes on the surface profile.  For a one-dimensional surface profile $z(x,t)$ with only surface diffusion, the continuity equation reads
\begin{equation}
{\partial z\over\partial t}+\Omega{\partial J\over\partial x}=0\, ,
\label{eq2}
\end{equation}
where $J$ is the surface current, which is given by the Nernst--Einstein equation:
\begin{equation}
J=-{nD\over k_BT}{\partial\mu\over\partial s}\, ,
\label{eq3}
\end{equation}
in which $n$ is the concentration of diffusing atoms, $D$ the surface diffusion constant, $k_B$ Boltzmann's constant, $T$ the absolute temperature, $\mu$ the excess chemical potential due to surface curvature, and $s$ the arc length along the surface.  Mullins' assumption (i) enables us to write the Gibbs--Thomson equation in a reduced form:~$\mu=\gamma\Omega\kappa$, where $\gamma$ is the isotropic surface free energy and $\kappa$ os the curvature of the surface. Under  Mullins' assumption (ii), we can neglect the difference between the arc length $s$ along the surface and the $x$-coordinate, and $\kappa\approx -z_{xx}$ (the negative sign accounts for the fact that the excess free energy is positive (resp.~negative) for convex (resp.~concave) regions).  Hence, by combining these approximations with (\ref{eq2}) and (\ref{eq3}), Mullins obtained the linear equation,
\begin{equation}
{\partial z\over\partial t}=-{nD\Omega^2\gamma\over k_BT}{\partial^4z\over\partial x^4}\, .
\label{eq4}
\end{equation}
This theory provides good agreement with experiment when the restrictions imposed by the two approximations are fulfilled.  But when, for example, facets form, the terms omitted in arriving at (\ref{eq4}) must be retained \cite{bonzel84,spohn93}.

\subsection{Morphological Evolution of Patterned Substrates}

Among the first systematic experimental studies of MOVPE on patterned substrates, a relevant one was reported in \cite{hersee86}, who examined growth as a function of temperature and time on V-grooves and mesas. The observed similarity of growth of Ga$_{1-x}$Al$_x$As for $0<x<0.5$ enabled the authors to use marker layers of different composition to follow the time  development of the layer structure. The growth rate was found to be facet- and temperature dependent, resulting in an increase of the ridge width and a  decrease of the width of the V-groove at low temperatures, with these trends reversing at high temperatures.  The authors proposed a four-step model for growth:~(i) The group-III precursor diffuses from the boundary layer of the gas flow to the heated substrate, where (ii) decomposition produces fragments containing the group-III atom, which (iii) diffuse (over a distance of microns), until (iv) incorporation.  Although \cite{hersee86} did not address the formation of quantum wires, the competition between facets and the emergence of new facets during growth were identified.

Subsequent work (e.g.~\cite{bhat88}) pointed out that not only are the facet-dependent growth rates material dependent, but that, at the center of the V-groove, a ``quantum-wire-like crescent shaped active region'' was observed, which opened the way to quantum wire lasers (for example, \cite{kapon89}). We will not review the extensive literature on this subject, but refer the interested reader to earlier reviews which comprehensively covered the phenomenology \cite{kapon99,wang06}. Instead we will briefly examine a subset of the early attempts to describe/model the ``special'' morphological evolution that MOVPE on patterned substrate delivered, mostly referring to V-groove quantum wires (as the pyramidal dot system appeared later).

We pause here to point out that a broad range of early experimental data was obtained in terms of growth temperature, reactor pressure, V/III ratio, etc. As will become clear below, some parameters have a major effect in determining the morphology, while others, such as the V/III ratio and the reactor pressure, exert a smaller influence. For this reason, no systematics have been collected in the literature, and these parameters have not been explored thoroughly. We will not address these two parameters in our review, but note that most of the work in the last 20 years was performed almost exclusively in low-pressure reactors. Moreover, the effective V/III ratio inside and outside the patterned area might not be the same because of facet-dependent precursor decomposition processes.

The competition between facets is an important point that merits further discussion.  Consider the following thought experiment. Suppose  two macroscopic wafers with different orientations, say a 1~cm$^2$ (100) and (111)A of GaAs, are placed  in an MBE chamber.  Adatom sticking coefficients determine incorporation probabilities, so if growth conditions are chosen so that desorption can be neglected, adatom diffusion lengths and incorporation rates will be different on the two wafers, but all adatoms will eventually incorporate, resulting in the same growth rate on the two non-competing wafers.  We can look at this another way by examining, say, a face-centered cubic crystal.  The site density of atomic positions on a given crystallographic plane scales inversely with the corresponding inter-planar spacing,  so for a fixed deposition rate (without desorption), the growth rate is independent of orientation.

On the other hand, if a single wafer is patterned to expose small (micron-size) contiguous facets, the different incorporation probabilities will result in facet competition:~long atom diffusion lengths will result in a higher adatom density on the facet with the highest sticking coefficient, regardless of the facet on which deposition initially occurred.  As a consequence, the actual local growth rate of each facet results from the competition between neighboring facets.

As revealed by several early studies, the morphological evolution during the epitaxy  \cite{madhukar83,stringfellow84}, and on patterned substrates in particular  \cite{ozdemir92,hersee86,guha89}, results from a complex interplay between the thermodynamic driving force towards equilibrium, and kinetics, which determine the overall rate at which the system relaxes toward equilibrium.  Thermodynamics includes surface energies of different facets (Sec.~\ref{sec3.1}), and bonding configurations of precursors, reaction fragments, and adatoms, while kinetics includes diffusion of surface species, precursor decomposition rates, both  of which are facet-dependent, and inter-facet migration \cite{hata90}.

\begin{figure}
\includegraphics[width=0.8\textwidth]{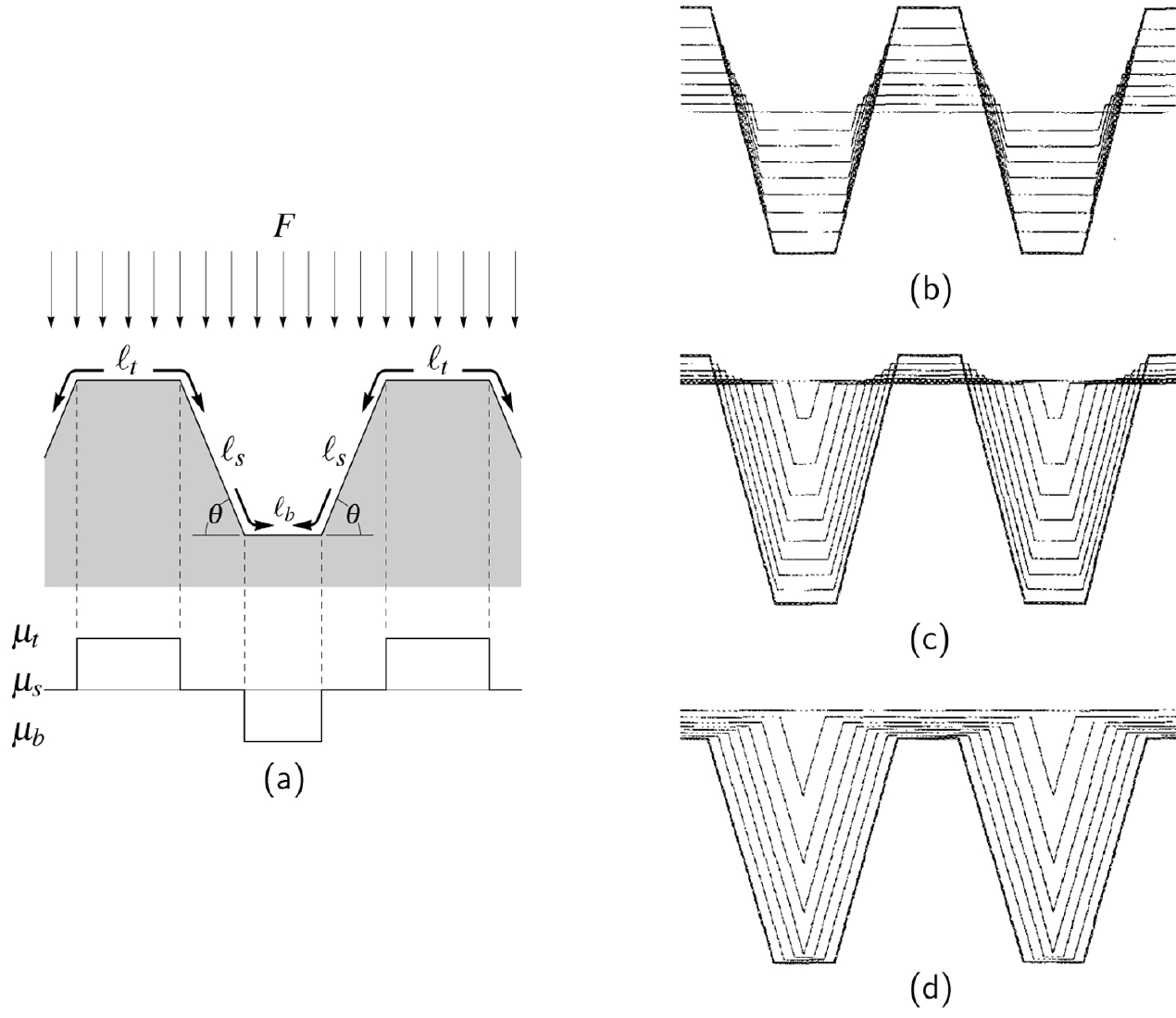}
\caption{(a) Section of a periodic sequence of mesas with top ($t$), side ($s$), and bottom ($b$) facets under the driving forces of an incident flux $F$ and equilibration due to capillarity, with the adatom currents indicated by the arrows along the surface.  The bottom panel shows the chemical potential across the surface. (b,c,d) Morphological evolution of $\sim5\mu$m facets with $F=0.5$~ML/s for a periodic sequence of mesas with facet lengths $\sim5\mu$m. The interfacet mass transfer rates at convex and concave corners is the same, and the attachment rate of adatoms to side facets is 10 times that of  the top and bottom facets. (b) capillarity dominates, (c) capillarity and growth are competitive, and (d) growth dominates. Successive profiles are not equally spaced in time, and planarization occurs in each case.\cite{ozdemir92}}
\label{fig3}       
\end{figure}

Ozdemir and Zangwill \cite{ozdemir92} reported the first systematic theoretical study of growth on patterned substrates that explicitly accounted for the competition between thermodynamics and kinetics. The authors formulated rate equations for the average adatom densities on each facet of a surface patterned with a periodic sequence of mesas composed of two inequivalent facets (Fig.~\ref{fig3}(a)).  This patterning produced a spatially-varying chemical potential which is the driving force for equilibration of the surface through capillarity. The chemical potentials across the surface is determined by applying (\ref{eq1}) to the top, side, and bottom facets, with result, respectively,
\begin{equation}
\mu_t=\mu_0+{2\Omega\gamma\over\ell_t}\, ,\qquad \mu_s=\mu_0\, ,\qquad \mu_b=\mu_0-{2\Omega\gamma\over\ell_b}\, ,
\label{eq5}
\end{equation}
where $\ell_t$ and $\ell_b$ are the widths of the top and bottom facets, respectively (Fig.~\ref{fig3}(a)), $\gamma=\gamma_s{\rm csc}\,\theta-\gamma_b{\rm cot}\,\theta$, and $\gamma_b=\gamma_t$, since these facets have the same orientation. The rate equations for the growth of each facet takes account of deposition, desorption, adatom attachment/detachment at kinks, interface mass transfer, and capillarity-driven mass transport.  Three regimes were studied: (i) capillarity dominates, (ii) capillarity and growth competitive, and (iii) growth dominates.  Planarization of the initial pattern was found always, but with the details of the evolution strongly dependent on the relative rates of the kinetic processes.  

Figure~\ref{fig3}(b,c,d) shows the morphological evolution of $\sim5\mu$m facets with a deposition of 0.5~monolayers (ML)/s for the three cases noted above when the interfacet mass transfer rates at convex and concave corners are the same and the attachment rate of adatoms to side facets is 10 times that of  the top and bottom facets.  Note, in particular that, if growth and capillarity are competitive (Fig.~\ref{fig3}(b)), the bottom facet width changes quite slowly, suggesting the possibility of a self-limiting width for other growth scenarios.  Although this model dealt only with atomic species, the formulation is flexible enough to allow inclusion of the group-III species.

We briefly mention attempts at describing the phenomenology of growth on patterned substrates during MOVPE as originating from gas-phase effects. The evidence relies on the agreement obtained for the diminished growth rate  observed in large buried patterned areas (typically tens of microns) from basic gas-phase diffusion models (see, for example, \cite{coronell91}). Precursor diffusion through the gas before reaching the substrate can be described by a concentration gradient supplied by a constant flux from gas-phase diffusion, and depleted by adatom incorporation into the substrate. As a consequence, higher positioned layers (e.g.~the broad top ridge) are exposed to a greater precursor concentration and greater adatom deposition rates, while the lower regions in the patterned area at a greater distance from the supply region,  exhibit a reduced growth rate. While such models reproduce the observed phenomenology in broad area patterning (tens or hundreds of microns), they fail for few-micron V-groove quantum-wire structures, not least because the V-groove growth rate in the bottom regions is often higher than the ridge region. 

\begin{figure}
\includegraphics[width=0.3\textwidth]{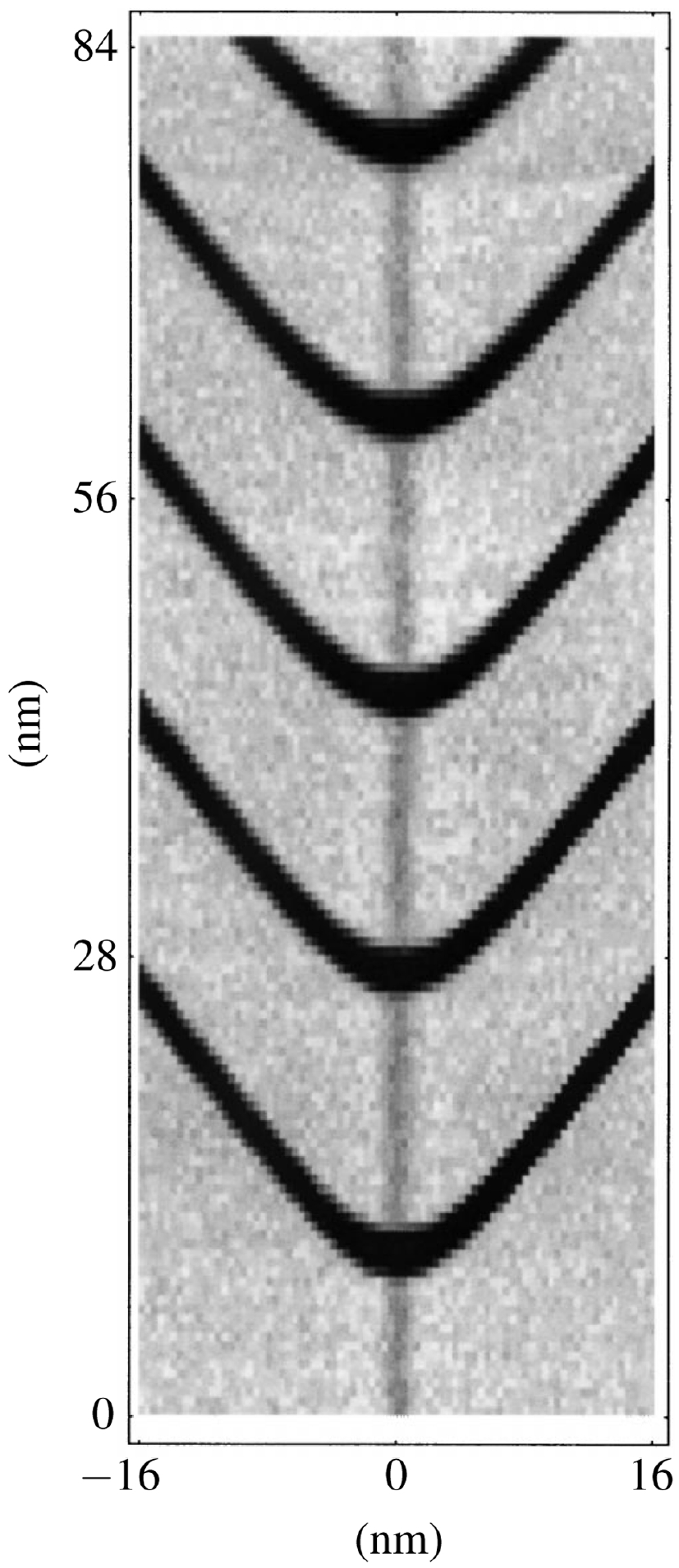}
\caption{From\cite{grosse00}. Cross-section of an array of Al$_{0.5}$Ga$_{0.5}$As/GaAs quantum wires obtained from KMC simulations using the zincblende structure with parameters optimized by fits to step flow experiments. The Al concentration is scaled between 0\% (black) and 60\% (white). The darker vertical stripe (corresponding to lower Al concentration) forms a vertical quantum well. Reprinted from Journal of Crystal Growth, \textbf{212}, Frank Grosse and Roland Zimmermann, \textit{Monte Carlo growth simulation for Al$_x$Ga$_{1-x}$As heteroepitaxy}, pp 128-137, Copyright (2000) with permission from Elsevier.}
\label{fig4}       
\end{figure}

An altogether different approach \cite{jones91} used the  Wulff construction and growth rates from different facets to determine the evolution of patterned GaAs substrates during MOVPE.  However, the neglect of the effect of adjacent facets on the growth rate of a given facet (cf.~\ref{eq1}), and the absence of alloying, which excludes segregation, pre-empts any consideration of quantum-wire formation in V-grooves.  Grosse and Zimmermann \cite{grosse00} studied quantum-wire and quantum-well formation in V-grooves with kinetic Monte Carlo (KMC) simulations of MBE with the zincblende structure of Al$_x$Ga$_{1-x}$As (Fig.~\ref{fig4}), but without explicitly accounting for the kinetics of As.  The parameters were optimized by comparison with  step-flow experiments, with the absence of As kinetics meaning that these parameters yield only {\it effective} rates.  This is common practise for growth under As-rich conditions, where the As kinetics are not rate limiting. The simulations reproduced the tendency toward crescent-shaped quantum wires near the apex of the V-groove, as well as the appearance of vertical quantum wells near the V-groove (Fig.~\ref{fig3}).  However, while the authors correctly identified the difference in Ga and Al as being responsible for the observed structures, the absence of any effects of precursors means that the length scales, time- and temperature-dependence do not correspond to those in V-groove quantum wires obtained by MOVPE.

\subsection{Self-Limiting Profiles of V-grooves during MOVPE}

An important step forward in our understanding about the morphological evolution of V-grooves during MOVPE came from  Biasiol and Kapon  \cite{biasiol99,biasiol02}, who analyzed extensive systematic experiments, including the dependence of the width of the bottom (100) base facet on temperature and AlGaAs alloy composition. This resulted in several fundamental advances:~(i) the identification of a ``self-limited profile'', i.e.~the steady-state shape of a specific layer and the conditions for its attainment, (ii) the characterization of facet-dependent growth rates during MOVPE and how they influence vicinal facet formation, and (iii) the analysis of the vertical quantum well associated with AlGaAs layers, i.e.~the segregation at the center of the V-groove of a Ga-rich quantum well and quantifying the Ga content as a function of alloy composition (and growth conditions) \cite{gustafsson95,biasiol96}. 

Based on these measurements, a model \cite{biasiol99,biasiol02} was proposed for AlGaAs growth by MOVPE in a V-groove, considering the different facets (Fig.~\ref{fig3}(a)) as different connected areas and imposing identical growth rates at facet boundaries. Apart from the explicit consideration of facets, this analysis is similar to that of Mullins (Sec.~\ref{sec3.1}).  The starting point is the continuity equation (\ref{eq1}) for the surface profile $z_i(x,t)$ for each facet $i$,
\begin{equation}
{\partial z_i\over\partial t}+\Omega{\partial J_i\over\partial x}=R_i\, ,
\label{eq6}
\end{equation}
For a V-groove (Fig.~\ref{fig3}(a)), $i=t,s,b$ for the top, side, and bottom facets, respectively.  The deposition flux $R_i=Rr_i$, where $R$ is the nominal  flux on a (001) planar reference surface, and $r_t=r_b$ because of the crystallographic equivalence of the top and bottom facets.  This anisotropy of the deposition onto different facets to account either for the presence of precursors in MOVPE, or shadowing effects in MBE. 

The adatom concentration on the side facets was determined from the steady-state continuity equation with the geometric and growth conditions yielding $n_s=Rr_s\tau_s/\Omega$, which is used for the density of adatoms at all facet boundaries in the Nernst--Einstein equation, where $\tau_s$ is the corresponding average incorporation time on the side facet. The capillarity fluxes are determined from discretized forms of the derivative in this equation and of the Nernst--Einstein equation (\ref{eq3}), yielding discrete second derivatives involving a given facet and the two adjacent facets.  Under these approximations, the continuity equations {\ref{eq6}) for the three facets are 
\begin{equation}
\begin{array}{rcl}
\displaystyle{{dz_t\over dt}}&=&\displaystyle{R_b+aR_s\biggl({\mu_s-\mu_t\over\ell_t^2}\biggr)\, ,}\\
\noalign{\vskip3pt}
\displaystyle{{dz_s\over dt}}&=&\displaystyle{R_s+{aR_s\over2\ell_s}\biggl({\mu_t-\mu_s\over\ell_t}-{\mu_s-\mu_b\over\ell_b}\biggr)\, ,}\\
\noalign{\vskip3pt}
\displaystyle{{dz_b\over dt}}&=&\displaystyle{R_b+aR_s\biggl({\mu_s-\mu_b\over\ell_b^2}\biggr)\, ,}
\end{array}
\label{eq8}
\end{equation}
in which $a=(2\lambda_s^2)/(k_BT)$, $\lambda_s=(D_s\tau_s)^{1/2}$ is the diffusion length on the side facet and the denominators in each discrete derivative represent the length over which curvature effects are negligible. 

\begin{figure}
\includegraphics[width=0.5\textwidth]{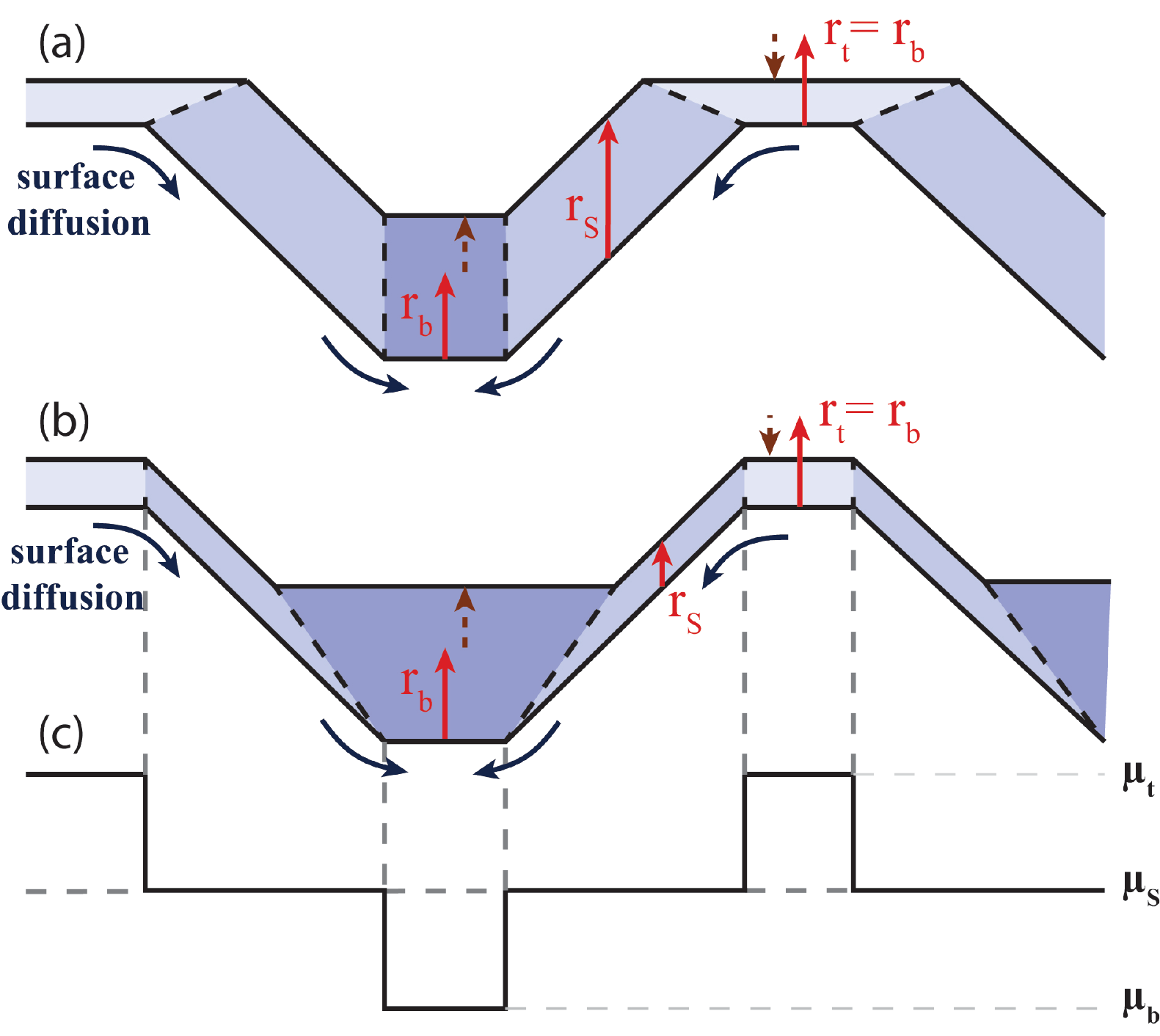}
\caption{Schematic profiles for a V-groove composed of three facets (not to scale). (a) Evolution of the growth front as commonly observed in MOVPE and (b) as in MBE. (c) Chemical potential at each facet.}
\label{fig5}       
\end{figure}

Self-limiting facet widths $\ell_t^\ast$ and $\ell_b^\ast$ of the top and bottom facets, respectively, are obtained when the growth rates on two neighboring facets are equal, with the results, assuming that $\ell_s\gg\ell_t$ and $\ell_s\gg\ell_b$,which implies
\begin{equation}
{dz_t\over dt}={d z_s\over dt}\approx R_s\, ,\qquad
{dz_b\over dt}={d z_s\over dt}\approx R_s\, ,
\label{eq9}
\end{equation}
whose solutions yield 
\begin{equation}
\ell_t^\ast=\biggl(-{a\Omega r_s\gamma\over\Delta r}\biggr)^{1\over3}\, ,\qquad
\ell_b^\ast=\biggl({a\Omega r_s\gamma\over\Delta r}\biggr)^{1\over3}\, ,
\label{eq10}
\end{equation}
where $\Delta r=r_s-r_t=r_s-r_b$.  These results are summarized in Fig.~\ref{fig5}. The evolution of the surface profile for $\Delta r>0$ is shown in Fig.~\ref{fig5}(a).  The additional adatom current due to capillarity at the bottom facet exactly balances the effect of $r_s$, yielding self-limiting growth at this facet.  At the top facet, however, capillarity further reduced the growth rate on this facet, which therefore expands and leads to the planarization of the surface.  This is similar to Fig.~\ref{fig3}(c), which shows the competing effects of capillarity and growth and is commonly observed in MOVPE.  Figure~\ref{fig5}(b) shows the surface profile for $\Delta r<0$. Capillarity can compensate for the growth rate anisotropy at the top facet, while the bottom facet will always grow faster than the sidewalls, thereby expanding and leading eventually to planarization.  This is the morphological evolution commonly seen in MBE.

Biasiol and Kapon extended their model to include two species to study the effect of Ga segregation on the self-limiting growth of Al$_x$Ga$_{1-x}$As as a function of $x$.  As AlAs-GaAs alloys are ideal solution over the entire range of concentrations \cite{panish72}, the chemical potentials for Al and Ga on each facet were first written as
\begin{equation}
\begin{array}{rcl}
\mu_i^{\rm Al}(x)&=&\mu_i+k_BT\ln x\, ,\\
\noalign{\vskip6pt}
\mu_i^{\rm Ga}(x)&=&\mu_i+k_BT\ln(1-x)\, ,
\end{array}
\label{eq11}
\end{equation}
where the superscripts refer to AlAs (Al) and GaAs (Ga),  $\mu_i$ are the chemical potentials of the top ($i=t$), side ($i=s$), and bottom ($i=b$) facets in (\ref{eq5}), and the additional terms result from the entropy of mixing of an ideal solution.  As the Al and Ga atoms can freely exchange in the alloy, the entropy is maximized by a random solution along each facet. Segregation of Ga at the bottom facet is enforced by setting the concentration $x_b$ there according to
\begin{equation}
x_b(x,k)={x\over  x(1-k)+k}\, ,
\label{eq12}
\end{equation}
where $x(k,0)=0$, $x(k,1)=1$ always, and $x_b(k,x)<x$ for $k>1$, with $k$ an adjustable parameter. Hence, the chemical potentials (\ref{eq11}) at the bottom facet are $\mu_b^{\rm Al}(x_b)$ and $\mu_b^{\rm Ga}(x_b)$. A procedure analogous to that leading to the system of equations (\ref{eq8}) produces partial growth rates for each component.  The self-limiting width is then determined from an alloy version of (\ref{eq9}),
\begin{equation}
x{dz_b^{\rm Al}\over dt}+(1-x){dz_b^{\rm Ga}\over dt}=
x{dz_s^{\rm Al}\over dt}+(1-x){dz_s^{\rm Ga}\over dt}\approx
xR_s^{\rm Al}+(1-x)R_s^{\rm Ga}\, ,
\label{eq13}
\end{equation}
which yields a cubic equation for $\ell_b^\ast$  whose solution reduces to the simple form (\ref{eq9}) only in the limits $x=0$ and $x=1$.

\begin{figure}
\includegraphics[width=0.7\textwidth]{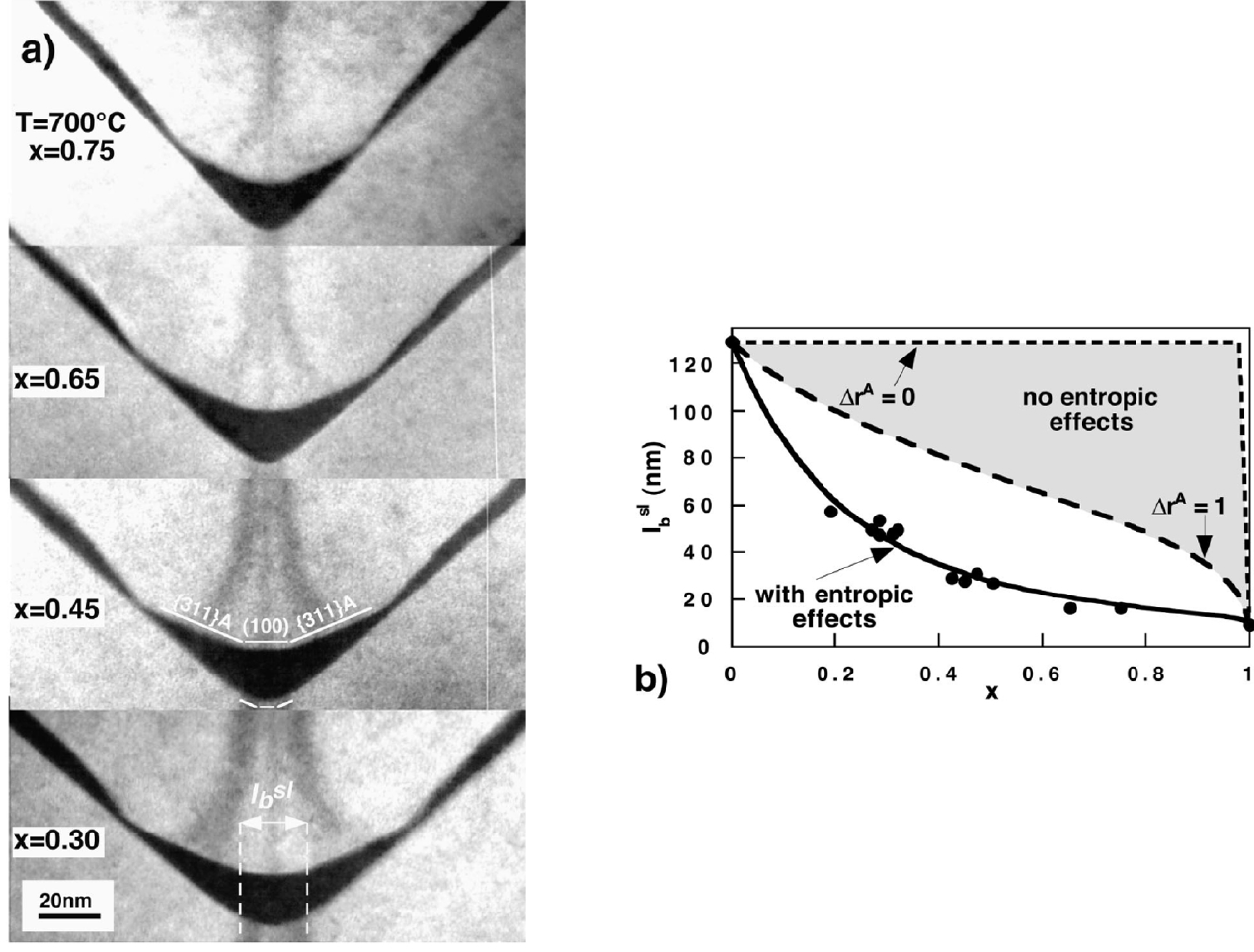}
\caption{From \cite{biasiol02}:(a) Cross-sectional TEM images of self-limiting Al$_x$Ga$_{1-x}$As alloys for the indicated compositions grown by low pressure OMCVD at 700$^\circ$C.  (b) Comparison of measure self-limiting widths with an optimized fit to the solution of (\ref{eq12}) as a function of $x$ at 700$^\circ$ (solid line).  The long- and short-dashed lines bounding the shaded region represent results obtained without the entropy of mixing and setting $\Delta r^{\rm Ga}=0.22$ and $\Delta r^{\rm Al}=1$ or $\Delta r^{\rm Al}\to0$, respectively. Reprinted figure with permission from G. Biasiol, A. Gustafsson, K. Leifer, and E. Kapon, Physical Review B, \textbf{65}, p 205306 (2002), Copyright (2002) by the American Physical Society.}
\label{fig6}       
\end{figure}

The results of this analysis are shown in Fig.~\ref{fig6} \cite{biasiol02}.  A series of TEM images is shown in Fig.~\ref{fig6}(a) width nominally 5-nm-thick GaAs markers inserted to measure the self-limiting Al$_{x}$Ga$_{1-x}$As profile widths. The images show that the bottom profile sharper as $x$ is increased, while the angle between the sidewalls decreases slightly.  The recovery of the self-limiting profile after broadening during GaAs growth is accompanied by the convergence of the three vertical quantum well branches from each bottom facet (cf.~Fig.~\ref{fig4}).  The dependence of the self-limiting width of the bottom profile on the composition for a series of samples grown at 700$^\circ$C is shown Fig.~\ref{fig6}(b). To within uncertainties of several experimental quantities, the  model provides a good fit to the measurements.  An important aspect of this fit and the comparisons in Fig.~\ref{fig5}(b) indicate that the entropy of mixing is an essential part of this agreement.

It is worth noting that the entropy of mixing terminology, necessary to this model, can be expressed as a simple Fick's law term. This is best understood in light of the origin of the formulation of the entropy on mixing when applied to the surface diffusion of adatoms, which can be found in \cite{tsao92}, for example. There, a two-components alloy ($a_{1-x} b_x$) condensed phase is considered and the entropy of mixing is calculated for $N_{\rm tot}=N_a+N_b$ lattice sites ($N_{\rm tot}$ is the total number of lattice sites $N_a$ is the number of type a atoms and $N_b$ is the number of type $b$ atoms) with no empty sites (i.e.~incident atoms always refill an empty site caused by incorporation). The mixing is given by the random exchange of the adatoms in the lattice following the diffusion process. In this hypothesis it is easy to show that the diffusion term arising from the mixing chemical potential can be described by Fick's law.
If we consider component $a$, we have for the chemical potential: 
\begin{equation}
\mu_a=k_BT\ln x=k_BTln\left( \frac{n_a}{n_a+n_b }\right) =k_BT\bigl[\ln n_a -\ln(n_a+n_b )\bigr]\, ,
\label{eq15}
\end{equation}
where $n_a=N_a/N$  is the surface density of type $a$ adatoms. We can calculate the surface current of $a$ adatoms using equation (\ref{eq3}):
\begin{equation}
J_a=-\frac{n_a D}{k_BT}  k_BT\left( \frac{1}{n_a}   \frac{\partial n_a}{\partial s}+\frac{1}{n_a+n_b}  \frac{\partial(n_a+n_b)}{\partial s}\right)
\label{eq16}
\end{equation}
Since this was calculated under the assumption of a completely filled lattice, the total density $N_{\rm tot}$ is constant and, therefore, the second derivative is zero, leading to
\begin{equation}
J_a=-D  \frac{\partial n_a}{\partial s}
\label{eq17}
\end{equation}
i.e.~Fick's law (the same applies for component $b$). The same result is obtained if, instead of considering a different species, $b$, we simply consider empty lattice sites, then the redistribution of $a$-type atoms on the free sites follows the same treatment. In general (considering, for example, more realistic scenarios in which, with time, deposition might result in the creation of morphologies with slight changes in the overall site availability for incorporation), entropy of mixing terms will fundamentally result in Fick's law under broader hypotheses than that discussed here, such as the case that the $N_{\rm tot}=N_a+N_b$ term is slowly varying compared to individual term variations. Regardless of the mathematical details, the entropy of mixing term in the model of Biasiol and Kapon plays an essential role, meaning that  a Fick's law-like diffusion current is needed to obtain agreement with experimental data, as we will see also in a different growth model in the following section. 

Nevertheless, despite the comparisons in Fig.~\ref{fig5}(b), the model is not without its drawbacks.  Although the expressions in (\ref{eq10}) establish the key role that growth rate anisotropy plays in the existence of self-limited profiles, the atomistic {\it origins} of this anisotropy is not included in the model.  In fact, growth rate anisotropies are best understood in terms of the decomposition  rate of the group-III precursors (typically, trimethylgallium and trimethylaluminum).  Although decomposition is a complex multi-reaction process, the overall decomposition rate can be described by a facet-dependent Arrhenius form.   But a more fundamental issue of the Biasiol--Kapon model is the treatment of segregation, which is built into the model through (\ref{eq12}), rather than being a consequence of the model. While the quality of the fit to experiment in Fig.~\ref{fig6}(b) indicates that the interplay between growth rate anisotropies of AlAs and GaAs and capillarity effects and corrections thereto can be parametrized by the model, that parametrization is likely to vary with growth conditions (partial growth rates and temperature) and the facet geometry (facet lengths and different etched patterns). In this sense, the predictive powers of this model are limited.

Moreover, subsequent attempts to adapt the model to the 3D case of the growth over a pyramidal recess did not succeed. We anticipate that the reason for this is to be found in the lack of a proper description of the decomposition anisotropy of the precursors, which in the pyramidal case is extreme, leading to the strong anisotropy of growth between (111)A and (111)B oriented facets.

%%%%%%%%%%%%%%%%%%%%%%%%%%%%%%%%%%%%%%%%%%%%%%%%%%

\section{Current Understanding and Models}
\label{sec4}

\subsection{Growth Rate Anisotropies in V-grooves and Inverted Pyramids}
\label{sec4.1}

Even before the work of Biasiol and Kapon \cite{biasiol99} was published, several authors had already pointed out that precursors decomposition anisotropy (i.e.~the fact that the decomposition process for a single precursor appeared to be crystallographic facet-dependent), was a major factor determining the V-groove profile (and wire formation) process. Here, we will mention the important contribution from Kaluza et al.~\cite{kaluza00}. The authors compared the self-limited profiles and relative facet growth rates as a function of MOVPE precursors.  Figure~\ref{fig7} shows striking evidence that the combination of trimethylgallium (TMGa) and trimethylaluminum (TMAl) provides strong anisotropies (i.e.~the lateral vicinal (111)A facets are growing significantly faster than the ridge (100) when the growth rate is measured along the growth direction) when compared to the combination TMGa/dimethylethylaminealane (DMEAAI), or even triethylgallium (TEGa)/ TMA, where no significant differences in the vertical growth rate can be seen.

\begin{figure}
\includegraphics[width=0.85\textwidth]{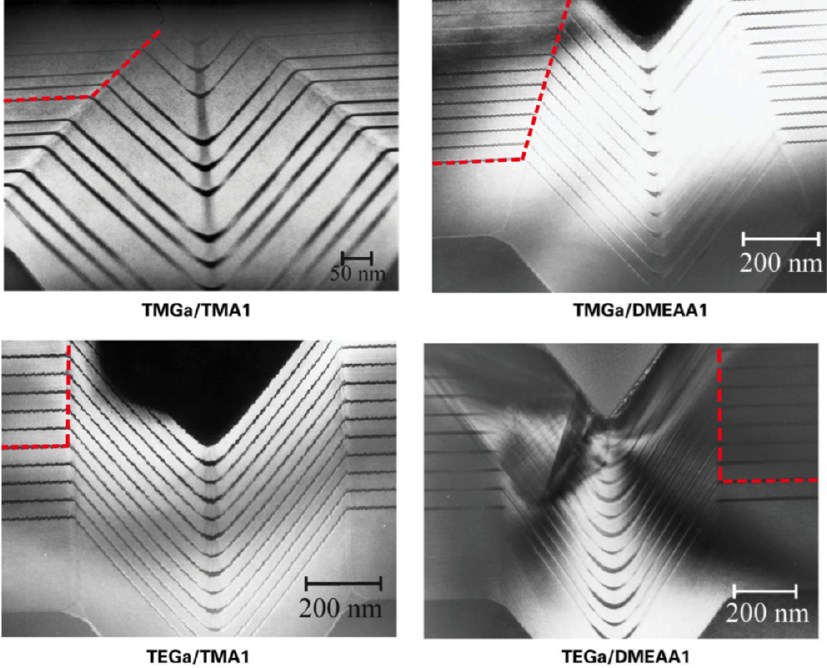}
\caption{TEM images of V-groove samples grown using different precursor combinations \cite{kaluza00}. (a) TMGa/TMAl, (b) TMGa/DMEAAl, (c) TEGa/TMAl and (d) TEGa/DMEAAl. Red dashed bars were added to the original image to highlight the difference in the evolution between the (100) and (111)A facets. Reprinted from Journal of Crystal Growth, 221, A. Kaluza, A. Schwarz, D. Gauer, H. Hardtdegen, N. Nastase, H. Luth, Th. Schapers, D. Meertens,A. Maciel, J. Ryan, E. O'Sullivan, On the choice of precursors for the MOVPE-growth of high-quality Al$_{0.30}$Ga$_{0.70}$As/GaAs v-groove quantum wires with large subband spacing, pp 91-97, Copyright (2000), with permission from Elsevier.}
\label{fig7}       
\end{figure}

These prominent differences suggested that the decomposition process could be facet dependent. Several years later Pelucchi {\it et al.}~\cite{pelucchi07} pointed out that, to simulate the unexpected behavior of pyramidal  quantum dot emission when a non-uniform pattern was utilized before growth, it is sufficient to assume that effectively no decomposition occurs on the flat (111)B surfaces, leaving the side (111)A surfaces providing the only decomposition sites. Indeed, by using a substrate pattern involving one pyramid placed at the center of a triangular (or hexagonal) area free of other pyramids, all embedded in a uniform array of pyramidal recesses, the emission wavelength of the isolated quantum dots showed that the isolated quantum dot was consistently thinner than the array of dots, i.e.~the emission was reproducibly blue shifted with respect to the quantum array emission wavelength (Fig.~\ref{fig8}). The model in Ref.~\cite{pelucchi07} was macroscopic, considering ``statistical'' diffusive components and dealing with pyramid to pyramid competition. In this sense, this does not impact the modelling of V-groove and pyramidal morphology covered in this review, but does show  that precursor decomposition is a major factor in nanostructure formation, and that proper growth modelling cannot be avoided. 

\begin{figure}
\includegraphics[width=0.65\textwidth]{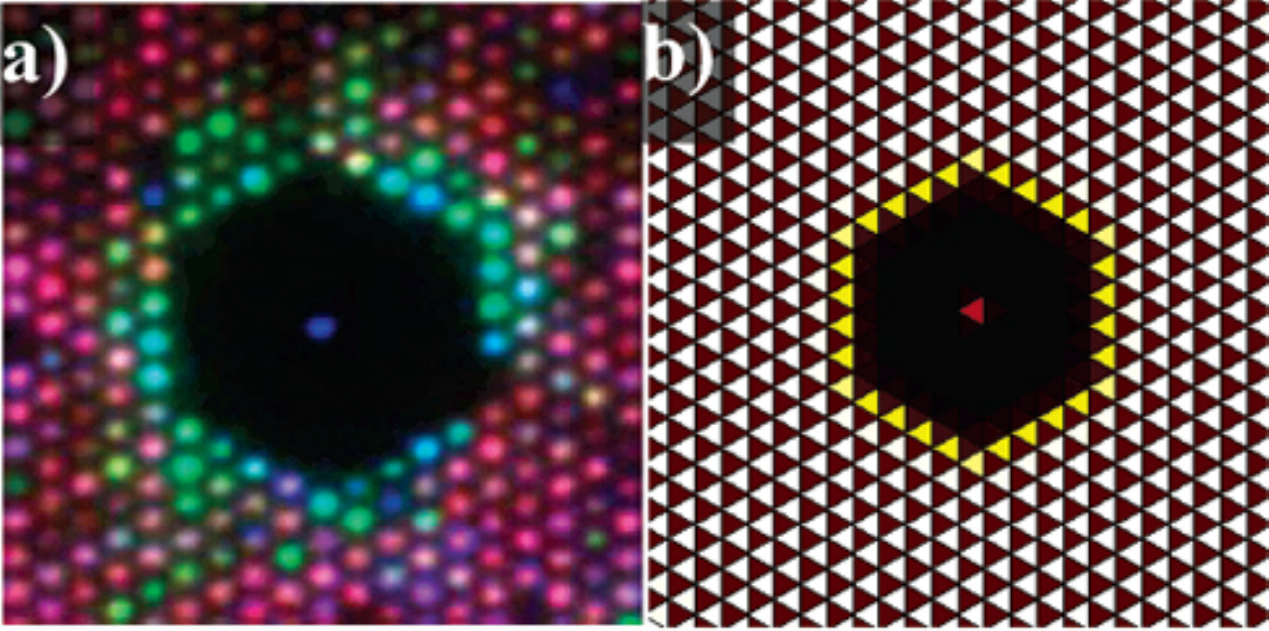}
\caption{From \cite{pelucchi07}: (a) false-color wavelength dispersive cathodoluminescence (CL) image of InGaAs QDs grown on a 500 nm pitch array with a hexagonal defect: the pyramid in the center of the defect presents a higher emission energy. (b) Simulation of the QD thickness distribution using the growth model in \cite{pelucchi07}. ¡°Hotter¡± colors represent thicker QDs: a good agreement with the CL image was found. Reprinted with permission from E. Pelucchi, S. Watanabe, K. Leifer, Q. Zhu, B. Dwir, P. De Los Rios, and E. Kapon, \textit{Mechanisms of Quantum Dot Energy Engineering by Metalorganic Vapor Phase Epitaxy on Patterned Nonplanar Substrates}, Nano Letters, \textbf{7}, pp 1282-1285 (2007). Copyright (2007) American Chemical Society.}
\label{fig8}       
\end{figure}

It is also obvious that this process takes place on a length scale comparable to the diffusion length for metalorganic precursors. This can be seen clearly in the pyramidal system, where fundamentally no decomposition happens on the top (111)B surfaces, and all adatom deposition appears on the lateral (111)A surfaces. Indeed, by growing on a substrate that is patterned with pyramidal recesses in a limited region only, it was observed \cite{pelucchi11} that the deposition on the (111)B surface takes place a few hundreds of microns away from the array of pyramids, where there is no competition between (111)A and (111)B surfaces.

\subsection{Reaction-Diffusion Equations with Growth-Rate Anisotropies}
\label{sec4.2}

The application of reaction-diffusion equations dates back to Ohtsuka and Miyazawa \cite{ohtsuka88}. who studied the evolution of one-dimensional patterns during MBE with a model that includes deposition, diffusion, and incorporation. Stepped surfaces were investigated, as were grooves and indentations, with qualitative agreement obtained for experiments on GaAs. But the most enduring aspect of this work is the recognition of the existence and importance of growth rate anisotropies for patterned substrates.  A later study \cite{ohtsuka99} extended the method to V-grooves and ridges.

The first model to implement the observations in Sec.~\ref{sec4.1} for V-groove quantum wire structure formation was reported in \cite{pelucchi11}.  The authors first provided experimental evidence of the facet dependent decomposition anisotropies.  As the decomposition process is thermally activated, facet-dependent growth rates should also be temperature dependent.  Figure~\ref{fig9} shows atomic force microscopy (AFM) images of a multilayer AlGaAs/GaAs structure grown by MOVPE at different temperatures.  Several differences between the nominally similar periods are seen. Here, we concentrate on the differences between the growth rates on the (100) ridge and on the (111)A vicinal planes. The evolution of boundaries between the two layers on the left side of the figure have been marked to highlight this effect. At low temperature, the growth rate on the (100) planes is minimal, with the entire growth process concentrated inside the V-groove. At the lowest temperatures this results in a significant lateral expansion of the (100) facet, which tends to quickly ``close'' (planarize) the patterned area. In this regime, the growth rate anisotropies are clearly evident. A sudden change in the planarization appears near $\sim$640$^\circ$C. The profile between the (100) ridge and the (111)A vicinal planes becomes steeper and the growth rate anisotropies reduce significantly, but still maintain a higher growth rate on the (111)A vicinal planes.

\begin{figure}
\includegraphics[width=0.85\textwidth]{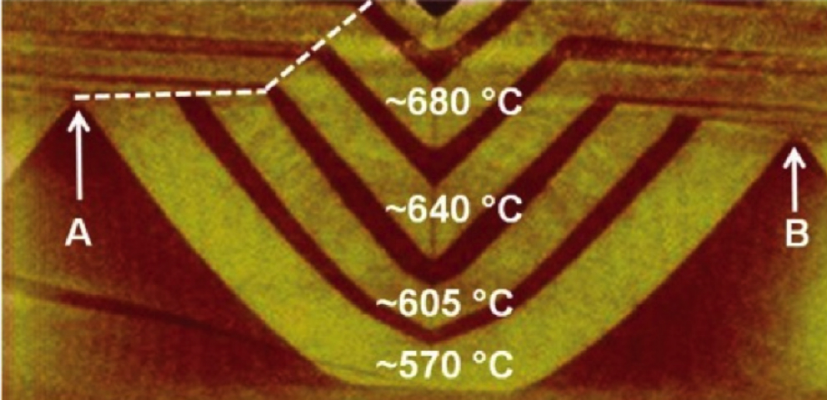}
\caption{AFM cross-sectional image of an AlGaAs/GaAs multilayer grown by MOVPE at the indicated temperature on a substrate with a 3~$\mu$m (from A to B) pitch pattern \cite{pelucchi11}. The dotted line indicates the evolution of the boundary between the (100) and (111)A planes. A thin oxide layer formed on the AlGaAs  layers after cleavage produces the difference  in the height signal of the AFM tip between AlGaAs and GaAs. Reprinted figure with permission from E. Pelucchi, V. Dimastrodonato, A. Rudra, K. Leifer, E. Kapon, L. Bethke, P. A. Zestanakis, and D. D. Vvedensky, Physical Review B, \textbf{83}, p 205409 (2011). Copyright (2011) by the American Physical Society.}
\label{fig9}       
\end{figure}

The model in \cite{pelucchi11} was of the type pioneered by Burton, Cabrera, and Frank \cite{burton51}, with the explicit inclusion of the contributions of decomposition rate anisotropies to growth rates, applied was  to the growth within  V-grooves.  Shortly thereafter Dimastrodonato {\it et al.}~\cite{dimas12} generalized thus approach to pyramidal quantum dots. The two models will be presented below from a unified perspective.

The growth model is based on a kinetics and comprises a set of reaction-diffusion equations. The key variable is the surface density of adatoms $n(x,t)$, which refers to a ``two-dimensional surface gas'' of the adatoms released after the decomposition of the precursors. In a stationary growth regime, the adatoms are generated at a fixed deposition rate ($F$), after which they diffuse according to Fick's first law with a diffusion coefficient ($D$) and are then incorporated onto the surface with an average lifetime ($\tau$) (i.e.~specific details apart, the model largely follows the approach of Burton, Cabrera, and Frank models that have appeared in the literature). All of these parameters are orientation-dependent, as different precursor decomposition rates, diffusion coefficients and incorporation lifetimes are assigned to each specific facet, resulting in a set of equations based on Fick's second law for each group-III species $k$ on facet $i$:
\begin{equation}
{\partial n_i^k\over\partial t}-D_i^k\nabla^2n_i^k=F_i^k-{n_i^k\over\tau_i^l}\, .
\label{eq14}
\end{equation}
For the growth of an alloy which is an ideal solution, the group-III species are independent from each other, resulting in a set of group III-species-dependent equations, while the group-V kinetics are disregarded, as high V/III flow ratios are generally used, so the group-V kinetics are not rate limiting. Despite the purely kinetic nature of the model, the diffusion coefficients and the lifetimes are treated as independent quantities implying that the diffusion current is an effective current of adatoms which takes into account the influence of thermodynamic effects (such as ``macroscopic'' capillarity terms). 

\begin{figure}
\includegraphics[width=0.8\textwidth]{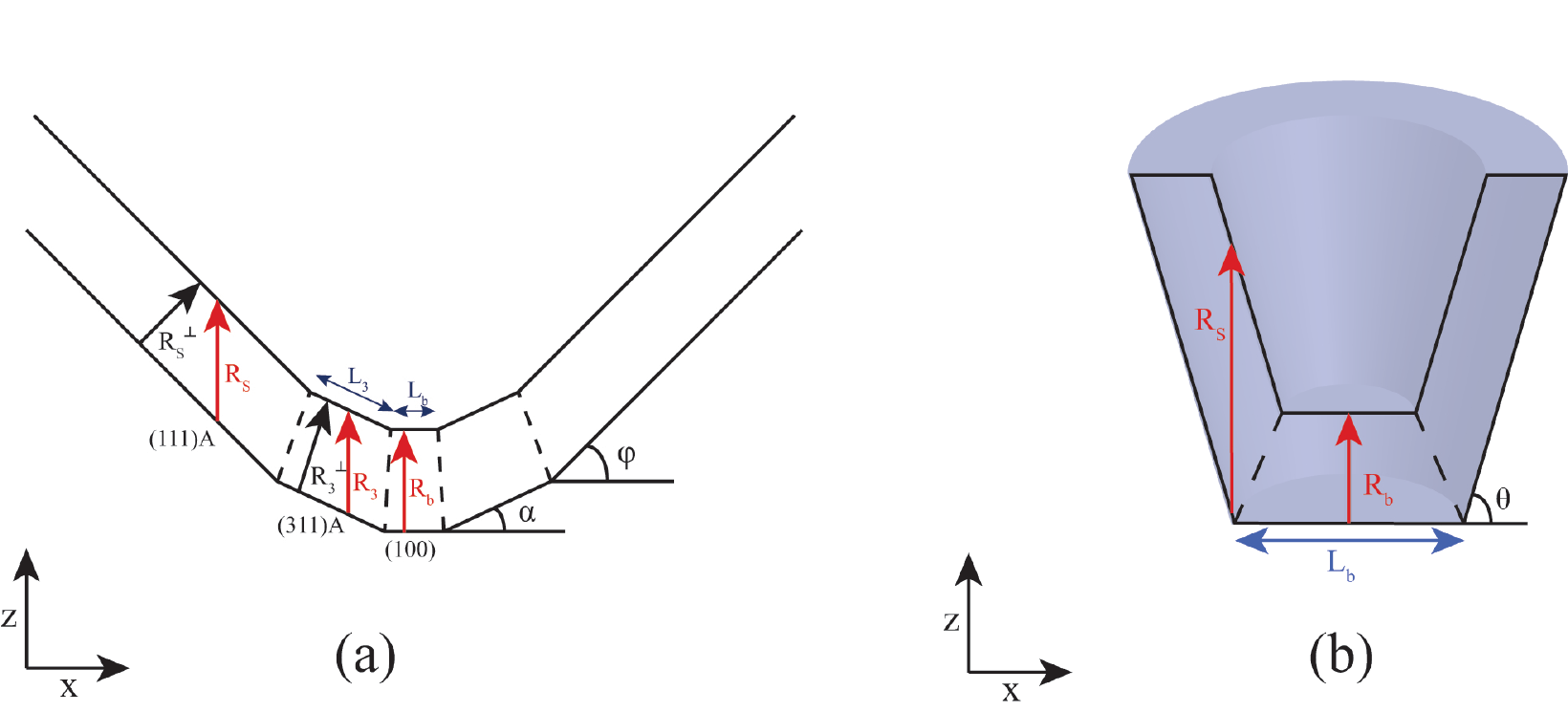}
\caption{(a) Schematic representation of the cross-section of a V-groove defining the surfaces to model the compositional and morphological evolution. The labels $b$, $s$, and 3 indicate the base facet, the lateral facets, and the intermediate (311)A facets, respectively. (b) Template showing growth rates, facets, and angles defined in the model for the conical approximation for pyramidal recesses. The labels $b$ and $s$ indicate, respectively, the base facet and the lateral facets.}
\label{fig10}       
\end{figure}

The specific geometry of the system (V-grooves or pyramidal recesses) comes into play when a general solution of the main equations (\ref{eq14}) is found by imposing continuity conditions for both $n_i^k$ and the resulting diffusion current: 
\begin{equation}
J_i^k=-D_i^k \nabla n_i^k\, .
\end{equation} 
This translates into a one-dimensional solution for the case of the V-groove, where the obvious symmetry for translations along the groove can be exploited (Fig.~\ref{fig10}(a)), while a more sophisticated approach needs to be found for pyramidal recesses. In the latter case a conical geometry was considered for simplicity (Fig.~\ref{fig10}(b)) and the simplified three-dimensional problem is solved analytically. Then, the resulting density $n_i^k$ determines the growth rate through
\begin{equation}
R_i^k={dz_i^k\over dt}={n_k^i\Omega\over\tau_i^k}\, .
\end{equation}
Finally, the overall growth rate must be the same on each facet if stationary growth conditions are assumed: 
\begin{equation}
\sum_k\bar{R}_i^k\Big|_{i=i_1}=\sum_k\bar{R}_i^k\Big|_{i=i_2}=\cdots\, ,
\end{equation}
where $\bar{R}i^k$ is the spatially-averaged growth rate on facet $i$ for a species $k$, from which both the equilibrium lateral dimensions of each of the facets composing the non-planar surface (therefore the self-limited profile) and the relative concentration of the group III species along each facet can be determined.

The general solutions of (\ref{eq14}) for the adatom densities on the facets forming the V-groove (Fig.~\ref{fig10}(a))  are
\begin{equation}
n_b^k(x)=F_b^k\tau_b^k+A_b^k\cosh\biggl({x\over\lambda_b^k}\biggr)\, ,
\end{equation}
for the (001) base facet, and
\begin{equation}
n_s^k(x)=F_s^k\tau_s^k+A_s^k\exp\biggl(-{x\over\lambda_s^k}\biggr)\, ,
\end{equation}
on the (111)A side facets. In these solutions, $A_i^k$ are arbitrary constants to be determined by boundary conditions between facets and $\lambda_i^k=(D_i^k\tau_i^k)^{1/2}$ is the diffusion length of species $k$ on facet $i$ prior to incorporation.

For the conical recess (Fig.~\ref{fig10}(b)), the corresponding solutions are
\begin{equation}
n_b^k(r)=F_b^k\tau_i^k+B_b^k I_0\biggl({r\over\lambda_b^k}\biggr)\, ,
\end{equation}
for the (111)B base facet, and
\begin{equation}
n_s^k(u^1)=F_s^k\tau_s^k+B_s^k K_0\biggl({\textstyle{1\over2}L_b+u^1\cot\theta\over \lambda_s^k\cos\theta}\biggr)\, ,
\end{equation}
for the (111)A side facet, where $I_0$ and $K_0$ are, respectively, the modified Bessel functions of the first and second kind of order zero, $r$ and $u^1$ are the non-Cartesian coordinates used to parametrize the surface, and $B_s^k$ and $B_s^k$ are arbitrary constants determined by the boundary conditions between the facets.

\begin{figure}
\includegraphics[width=0.9\textwidth]{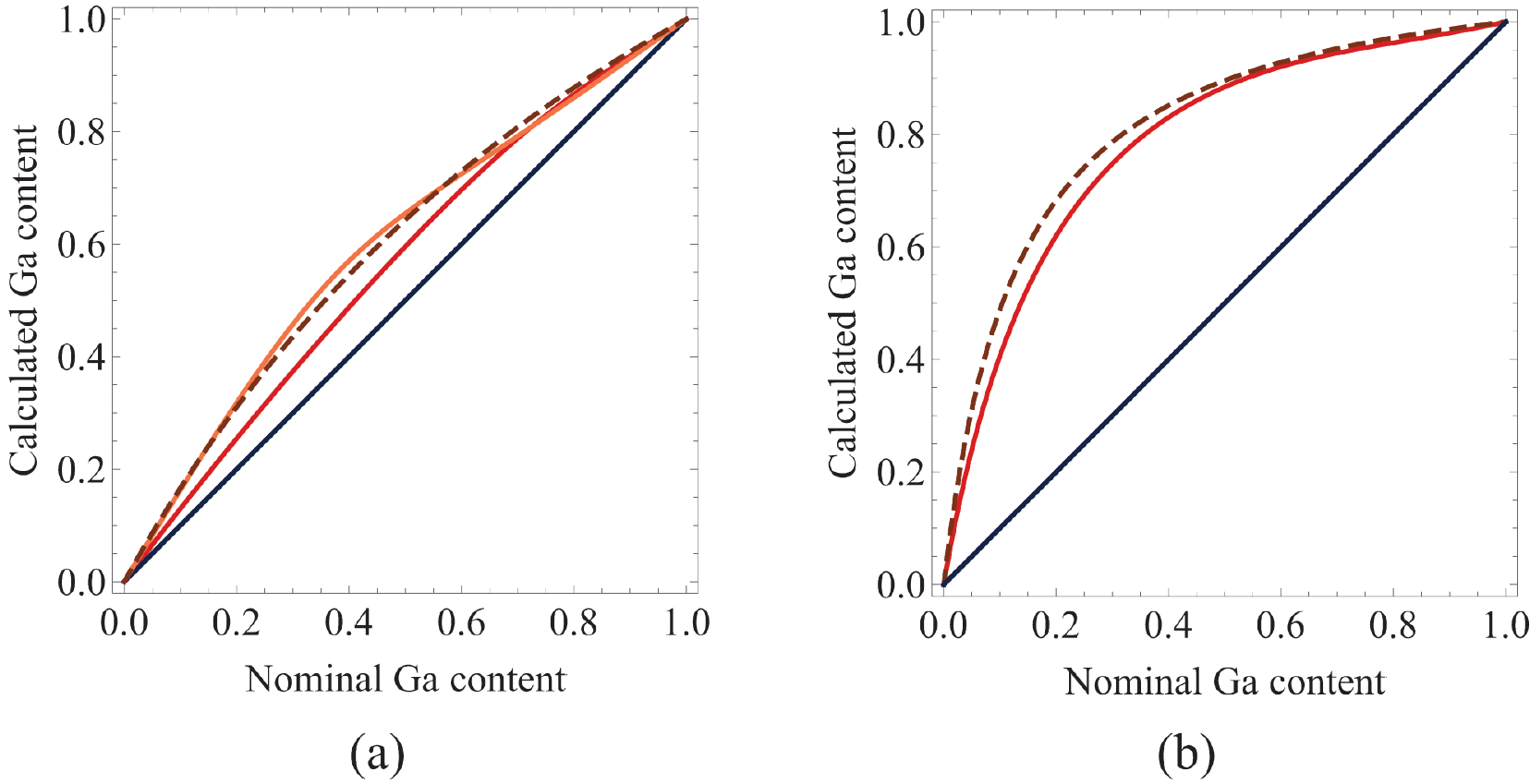}
\caption{(a) Calculated Ga relative growth rate in the middle of the (001) bottom facet of a V-groove for the model without (orange trace) and with the (311)A bottom facets (red trace). The blue line reproduces the nominal concentration and the red, dashed trace fits the experimental values of Ga concentration according to the experimental formula from Ref.~\cite{dimas13}. (b) Comparison between calculated steady-state (red trace) and fit to experimental values (red, dashed trace), from Ref.~\cite{dimas12}, of the Ga content in the pyramidal recesses as a function of the nominal alloy composition. The blue line reproduces the nominal concentration.}
\label{fig11}       
\end{figure}

Optimized fits of the foregoing solutions produced striking agreement with experimental measurements. For example, Fig.~\ref{fig11}, shows the vertical quantum well/wire (in the case of V-grooves and pyramidal quantum dots, respectively) experimental segregation {\it versus} theoretical prediction at a given growth rate. Similar agreement can be found for the self-limited profiles, which, should be said, can be matched to the theoretical work with a broad range of parameters, while the spatial dependence of segregation on the bottom (100) facet is by far more difficult to predict correctly. Indeed, the V-groove solution for Ga segregation in the vertical quantum well did not match experimental findings if a simplified model without the bottom (311) facet was attempted.  The inclusion of the (311) faceting and evolution was indeed necessary to reproduce the Ga segregation. On the other hand, the self-limited profile did not offer such a challenge.  These results are not to be considered as simple ``plain'' fittings.  All parameters appearing in the equations are known to have a well-defined temperature dependence. In Ref.~\cite{dimas12}, this is exploited by comparing experimental results for the self-limited profile as a function of growth, obtaining a good agreement between predictions and experimental results, with virtually no free fitting parameter. 

The overall model was shown to be indeed capable of describing a broader phenomenology than the one merely fitted. In Ref.~\cite{dimas13} the authors show that the experimentally verified change of vicinal (111)A surfaces in V-grooves (i.e.~change in the facet angle with respect to the growth direction) which appeared as a puzzle to the scientific community, was indeed a simple consequence of the model and the difference in growth rate between the bottom and top (100) surfaces.

These results  are based on the AlGaAs/GaAs system. Recently Moroni {\it et al.}~\cite{moroni15} showed that the model appears equally capable of describing In segregation when InGaAs V-groove quantum wires or pyramidal dots are considered, not only describing accurately previous experimental findings (Fig.~\ref{fig12}(a)), but also explaining a puzzling experimental result in the pyramidal system. As shown in Fig.~\ref{fig12}(b), lateral wires blue shift with increasing temperature, while the dots red shift. This was attributed by the model to different dominating factors:~for the dot the In segregation does not have a significant temperature evolution, so that the emission is dominated by the change in the self-limited profile of the GaAs barriers (which grows with temperature). In the case of the lateral wires, the self-limited profile is already significant in the temperature range considered and has a minor effect, while the In segregation/content in the lateral quantum wires has a small but measurable tendency to reduce with temperature, giving the observed blueshift.

\begin{figure}
\includegraphics[width=\textwidth]{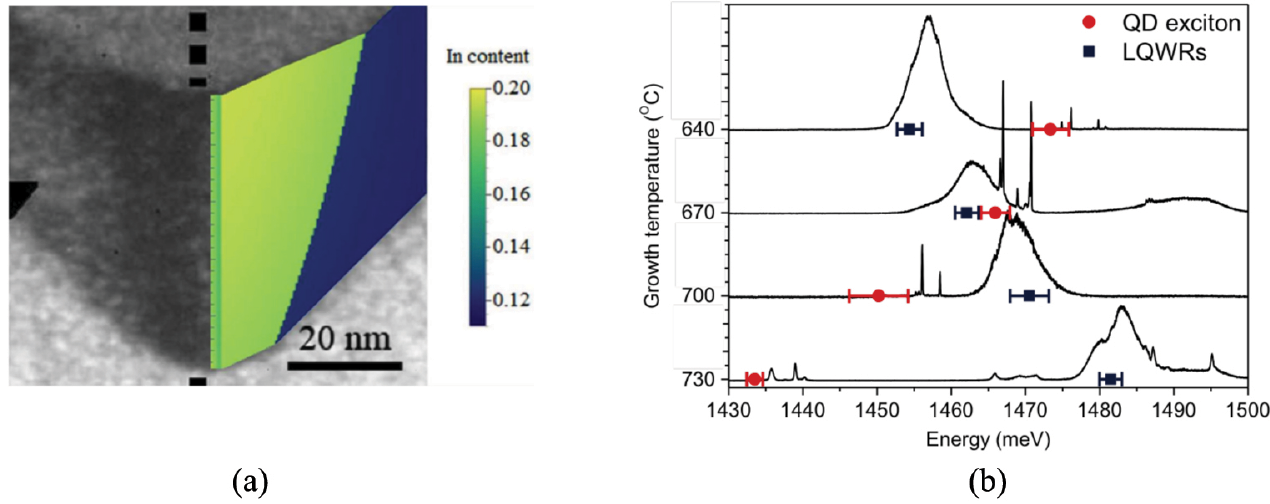}
\caption{(a) Comparison between the transient evolution of the experimental InGaAs V-grooved quantum wire (gray-scale cross-sectional TEM image, on the left) \cite{lelarge99}, and the profile resulting from the simulation (green and blue tone, on the right) \cite{moroni15}. (b) Photoluminescence spectra of four pyramidal quantum dot samples grown at different temperatures from \cite{moroni15}. The graph shows four representative spectra in which the emission of lateral quantum wires and quantum dots is seen to anti-cross as the growth temperature is changed, the marked points represent the typical energies resulting from a large statistics on the same samples. Reprinted from Stefano T. Moroni, Valeria Dimastrodonato, Tung-Hsun Chung, Gediminas Juska, Agnieszka Gocalinska, \textit{Indium segregation during III-V quantum wire and quantum dot formation on patterned substrates}, Journal of Applied Physics, \textbf{117}, p 164313 (2015), with the permission of AIP Publishing.}
\label{fig12}       
\end{figure}

\begin{figure}
\includegraphics[width=0.85\textwidth]{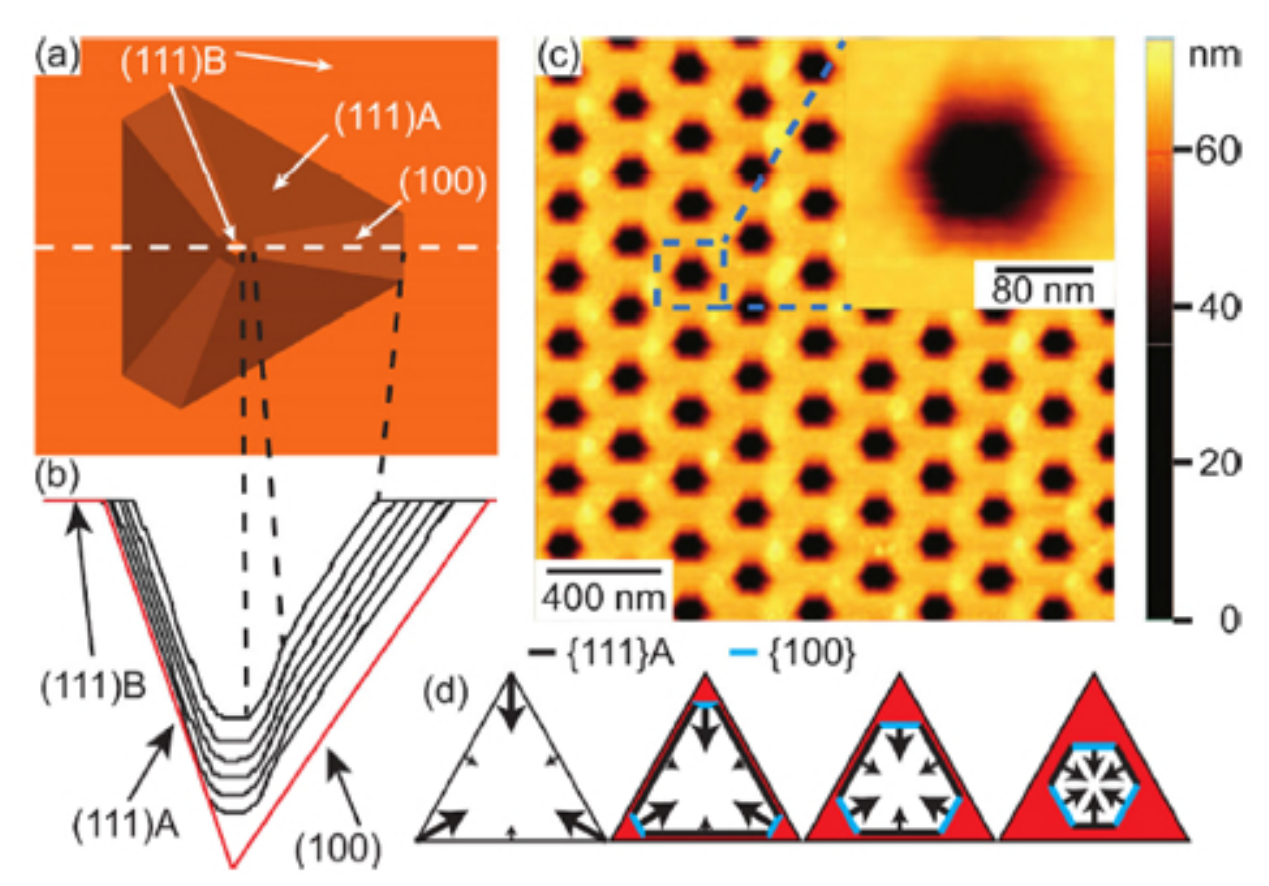}
\caption{From \cite{surrente16}: (a) 3D sketch of a pyramid after the development of a hexagonal bottom compared with (b) simulated cross section of an overgrown pyramid. The black lines represent the evolution of the profile during the deposition over the (red) original one. (c) $2\mu m\times 2\mu m$ AFM scan (topography signal) of an array of pyramids with 200 nm side and 250 nm pitch size after the growth of a thick (tGaAs = 8 nm) GaAs layer, showing the regular hexagonal shape of the pits. Inset: cropped image of a single recess. (d) Schematic illustration of the development of the hexagonal surface topology. The arrow thickness indicates the local growth rate. \{111\}A planes are depicted in black and \{100\} planes are depicted in blue. Reprinted from Nano Research, \textit{Self-formation of hexagonal nanotemplates for growth of pyramidal quantum dots by metalorganic vapor phase epitaxy on patterned substrates}, \textbf{9}, (2016), pp 3279-3290, Alessandro Surrente, Romain Carron, Pascal Gallo, Alok Rudra, Benjamin Dwir and Eli Kapon,  With permission of Springer}
\label{fig13}       
\end{figure}

In the same manuscript an important experimental observation was reported. While until then all models assumed a simple (111)A/(111)B/(111)A structure at the pyramidal center, Moroni {\it et al.}~\cite{moroni15} showed that a more complex faceting at the bottom of pyramidal recesses appears and is indeed necessary to properly link the bottom and lateral facets maintaining crystallographic continuity. This observation is shared in \cite{surrente16}, where for the first time a kinetic Monte Carlo simulation of GaAs growth inside pyramidal recesses is attempted. Despite the simplicity of the model implemented the authors report  a striking similarity between simulation and experimental morphology, correctly predicting the evolution not only of the center (111)B facet, including the extra high order faceting (see Fig.~\ref{fig13}), but also the lateral wire (100) expansion and formation.

\section{Summary and Outlook}
\label{sec5}
The development of our understanding of nanostructure formation on patterned substrates during MOVPE has been driven by the availability of systematic experiments of the growth of quantum wires within V-grooves and quantum dots within inverted pyramids.   The accompanying developments in the theory and modelling of these processes has seen increasingly refined descriptions of the fundamental kinetic processes and their consequences for the spatial distributions of alloy concentrations.  The most detailed of such approaches are based on  a three-step model:~(i) the arrival of polyatomic precursors onto a heated substrate, followed by the diffusion and decomposition of these precursors, releasing the atomic constituents of the growing material, and (iii) the diffusion and corporation of these atoms. The pronounced facet-dependence of (ii) and (iii) is essential for understanding why MOVPE is suitable for forming ordered nanostructures on patterned substrates and must be included in any quantitative model of nanostructure formation.

Typical length scales on patterned substrates are measured in microns, so continuum formulation of growth kinetics have dominated the modelling landscape.  But the effectiveness of the continuum picture (Sec.~\ref{sec4.2}) has provided the impetus for using kinetic Monte Carlo simulations of models with atomic-scale resolution to tackle with stronger accuracy the complexity of the full growth process. While KMC simulations have appeared sporadically over the years \cite{grosse00,haider94,koshiba95,lelarge99a,esen14}, the recent report in \cite{surrente16} illustrates both the power of this methodology, but also the limitations of conventional applications with regard to system sizes, as this study (and all others cited) are limited to sub-micron structures.  Alternative strategies are available, however, including parallelization based on spatial domain decomposition \cite{haider95} and hybrid schemes that incorporate a continuum description of diffusion \cite{schulze03}.

Large-scale simulations should enable descriptions with higher accuracy of the pyramidal growth process, allowing the effective engineering of nanostructure formation for a required application, along the lines of a recent report \cite{chung16} where vertical quantum wire structures were engineered for selective electric carrier injection into a single entangled photon emitter. This should help in implementing large-scale arrays of identical quantum emitters with a powerful impact on quantum technology roadmaps.

\section*{Acknowledgements}
This research was enabled by Science Foundation Ireland under grants SFI/05/IN.1/I25, 10/IN.1/I3000, 15/IA/2864.

\end{document}